\documentclass[singlespacing,12pt,onecolumn]{article}
\pdfoutput=1
\usepackage{graphicx,anysize,float,wrapfig}
\usepackage[square,sort&compress,super,comma,numbers]{natbib}
\usepackage{natbibspacing}
\marginsize{1in}{1in}{1in}{1in}
\usepackage{bm,epstopdf,bbm,multirow,epsfig,subfigure}
\usepackage{amsmath,amssymb,mathrsfs}
\usepackage{epic,eepic,setspace}
\usepackage{makeidx}
\usepackage{hyperref,varioref}

\begin{document}
\bibliographystyle{abbrvnat}

\noindent \textsf{Appeared in Macromolecular Theory and Simulations, {\bf 20}, 446(2011)} \smallskip \\
\noindent
 DOI : {\it 10.1002/mats.201100002}  \bigskip \\
 Article Type: Feature Articles/Reviews \bigskip \\
\noindent
\begin{center}
\textbf{\Large Modeling anisotropic elasticity of fluid membranes}\medskip \bigskip \\
\noindent {\large N. Ramakrishnan$^{1, *}$, P. B. Sunil Kumar$^{1,\dagger}$} \medskip \\
\noindent \textit{$^{1}$Department of Physics, Indian Institute of Technology Madras, Chennai, 600036, India} \\
\texttt{Email : $^{*}$ram@physics.iitm.ac.in, $^{\dagger}$sunil@physics.iitm.ac.in }\bigskip \\

\noindent{\large John H. Ipsen$^{2}$} \medskip \\
\noindent \textit{$^{2}$MEMPHYS- Center for Biomembrane Physics, Department of Physics and Chemistry,}\smallskip \\
\textit{University of Southern Denmark, Campusvej 55, DK-5230 Odense M, Denmark} \smallskip\\
\texttt{Email : ipsen@memphys.sdu.dk }\bigskip \\ 

\end{center}

\noindent The biological membrane, which compartmentalizes the cell and its organelles,
\begin{wrapfigure}{r}{3.0cm}
\begin{center}
\includegraphics[scale=0.35]{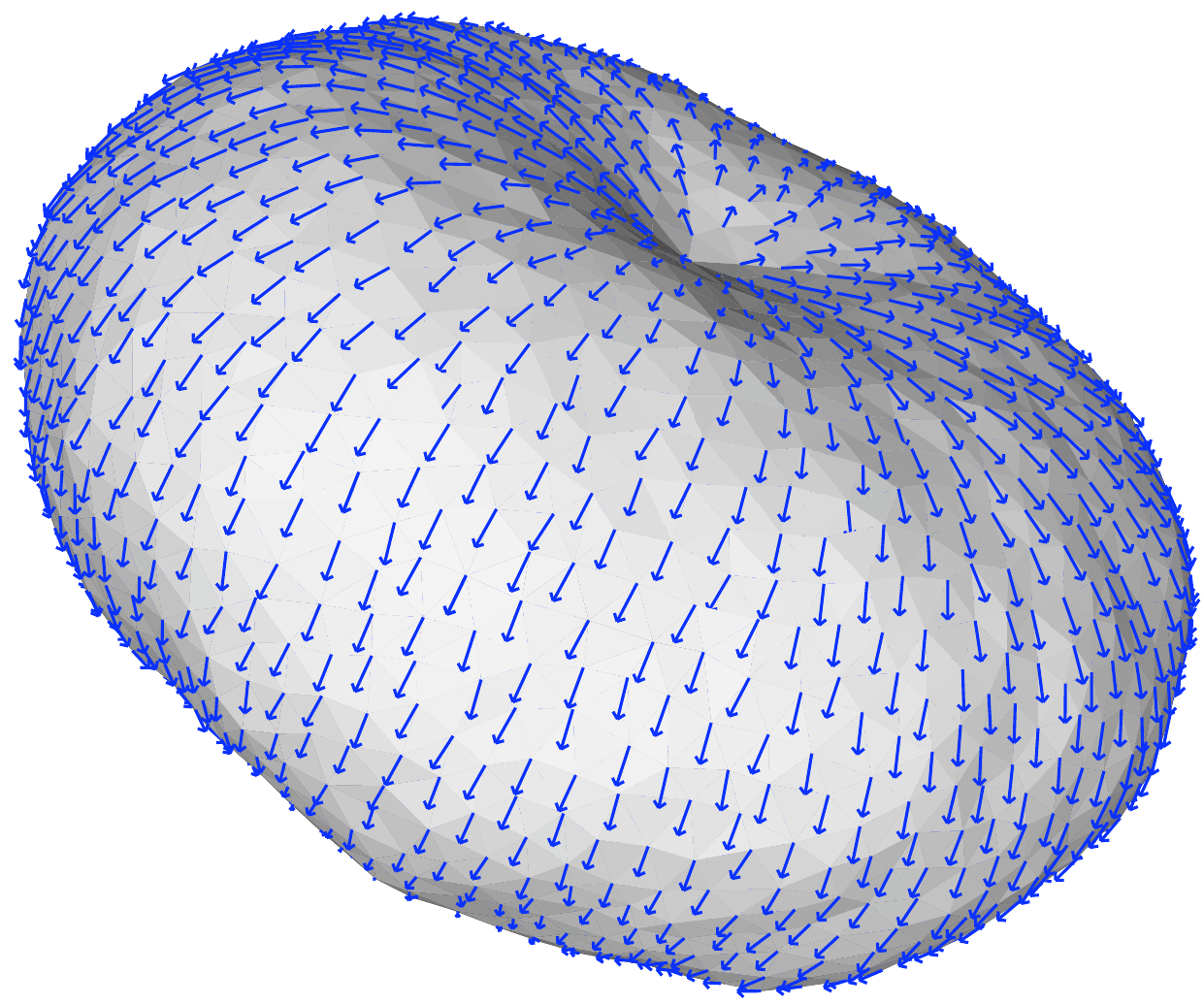}
\end{center}
\end{wrapfigure}
exhibit wide variety of macroscopic shapes of varying morphology and topology.s
 A systematic understanding of the relation of  membrane shapes to composition, external field, environmental conditions etc. have important biological relevance. Here we  review  the triangulated surface model, used in the macroscopic simulation of membranes and the associated Monte Carlo (DTMC) methods.  New techniques to calculate surface quantifiers, that will facilitate the study of additional in-plane orientational degrees of freedom, has been introduced. The mere presence of a polar and nematic fields in the ordered phase drives the ground state conformations of the membrane to a cylinder and tetrahedron respectively. \\

\section*{Introduction} \label{intro}
The biomembrane, that define the shapes of the cells and organelles that they enclose, are a few nanometer thick, with their lateral size extending from ten's of nanometers up to microns~\citep{Alberts:1994} . These  membranes can organize itself into a wide array of shapes, from relatively smooth shapes, as in plasma membranes, to complex cisternae, as in the ER and Golgi.   Understanding the properties of membranes and their interaction with other macromolecules, at multiple scales, is absolutely essential in modeling most biological processes. However, the complexity arising from the chemical diversity of membrane components hinders experimental investigations and modeling at the molecular scale. In the classical model proposed by Canham$^{\cite{Canham:1970p61}}$ and Helfrich$^{\cite{Helfrich:1973p693}}$, the homogeneous membrane, with its components having a diffusive degree of freedom, represents a two dimensional fluid surface with conformational energy,
\begin{equation}
\label{eq:Hsurf}
\mathscr{H}_{sur}=\int_{S} d{\mathbf{S}}\,\left \{ \frac{\kappa}{2} H^{2}+\sigma+\kappa_{G}R \right \}+\int_{V} \Delta p  ~ d{\mathbf{V}}.
\end{equation}
The phenomenological parameters, $\kappa$, $\sigma$, $\kappa_{G}$ and  $\Delta p$, respectively the bending rigidity, surface tension, Gaussian rigidity and osmotic pressure difference, depend on the chemical composition of the membrane. $H$ and $R$ are the mean and Gaussian curvature of the surface.  For an excellent review  on the applicability of this and other related models  see the article by Seifert.$^{\cite{Seifert:1997p1058}}$ \\

  The anisotropy in bending modulus, arising from the presence of embedding proteins  and nanostructured membrane domains, have been found to influence membrane shapes. For instance, the presence of surface membrane proteins containing BAR domains, like Nexin, Dynamin, Caveolin etc.,  have been shown to induce a wide array of shapes to the membrane.$^{\cite{McMahon:2005p274,Zimmerberg:2006p510,Dawson:2006p113,Voeltz:2007p1399}}$ Coatomers, Dynamin  and F-BAR domain containing proteins   stabilize {\it tubular} membrane structures, while proteins like Caveolin, I-BAR and some class of toxins like the Shiga toxin$^{\cite{Romer:2007p535}}$ form {\it caveola}, that are tubular structures drawn into cytoplasm.  Protein induced membrane curvature  is believed to be one of the possible mechanisms controlling  membrane morphology. This idea is further supported by molecular simulations where proteins oriented in an ordered phase, have been shown to induce a spontaneous curvature to the membrane.$^{\cite{Arkhipov:2008p257,Yin:2009p255,Blood:2006p358}}$ The protein distribution  and their cooperative effect on membrane morphology has been theoretically investigated by extending the existing membrane models.$^{\cite{Fournier:1996p488,Fournier:1998p00,Biscari:2006p128,Frank:2008p1047}}$ However, the realm of these theoretical models are restricted to axisymmetric membrane shapes and small deviations around it. 

\section*{Triangulated surface model for membrane} \label{sec:triang} 

\begin{figure}[!h]
\centering
\subfigure[]{
\begin{minipage}[]{0.4\textwidth}
\vspace{10pt}
\includegraphics[width=5cm]{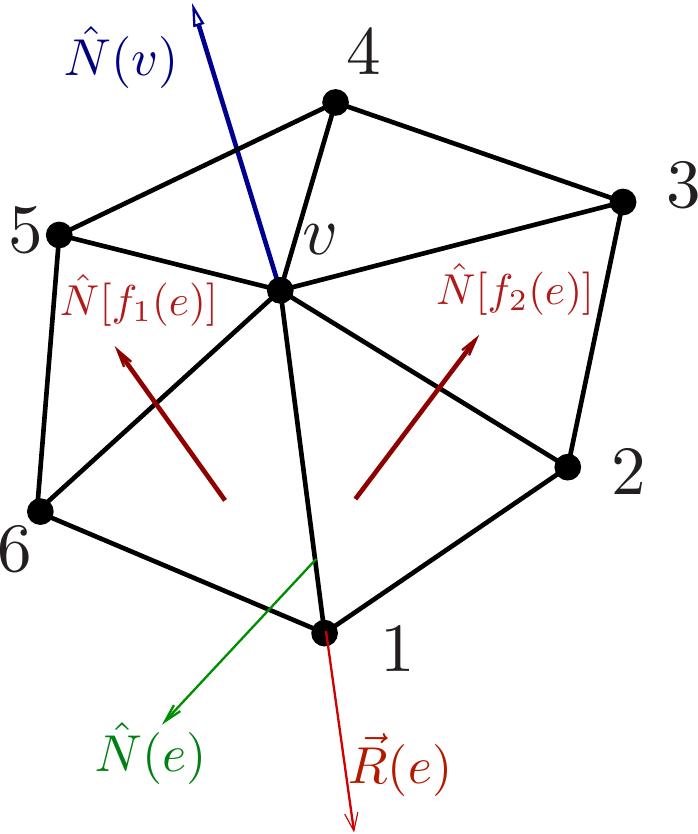}
\label{fig:patch}
\end{minipage}
}
\subfigure[]{
\begin{minipage}[]{0.40\textwidth}
\vspace{40pt}
\includegraphics[width=6cm]{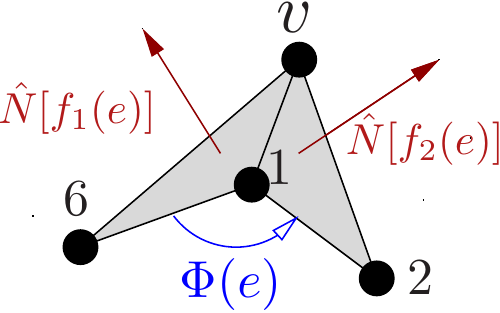}
\vspace{30pt}
\label{fig:dihedral}
\end{minipage}
}
\caption{(a) A surface patch showing a one ring neighbourhood around vertex $v$. $\hat{N}(v)$ represents the normal to the tangent plane at $v$. $\vec{R}(e)$ represents the vector along an edge $e$, while $\hat{N}(e)$ is its normal. (b) the signed dihedral angle $\Phi(e)$ between faces, $f_{1}$ and $f_{2}$, sharing an edge $e$. }
\end{figure}

In this approach, the vesicle, a two dimensional closed surface, embedded in three dimensional space, is discretized into a set of, interconnected, self avoiding, triangular plaquettes. Each of these plaquette correspond to a flat bilayer patch of the membrane. The $T$ plaquettes, constituting $L$ independent links, intersect at $N$ vertex points. The topology of the closed discretized surface, is defined  by the Euler characteristics $\chi=N+T-L$.

The implicit discretization of the elastic energy of the membrane in Equation\eqref{eq:Hsurf} is given by
\begin{equation}
\label{eq:Gompdisc}
\mathscr{H}=\lambda_{b} \sum_{v} \sum_{\{e\}_{v}} \left \{1-\hat{N}[f_{1}(e)] .\hat{N}[f_{2}(e)]\right\}+\Delta p \ V.
\end{equation} 

As illustrated in Figure \ref{fig:patch}, $\{e(v)\}$ denote the set of links, a vertex $v$ makes in its one ring neighbourhood. $\hat{N}[f_{1}(e)]$ and $\hat{N}[f_{2}(e)]$ are respectively the normals to faces $f_{1}$ and $f_{2}$ sharing an edge $e$. $V$ is the volume enclosed by the surface. The bending rigidities in Equations\eqref{eq:Gompdisc} and \eqref{eq:Hsurf} are related to each other as, $\lambda_{b}=\sqrt{3}\kappa$ for a sphere and   $\lambda_{b}=2 \kappa/\sqrt{3}$ for a cylinder.\cite{Piran:2003}  The triangulated mesh can be evolved using various techniques suiting the nature of study. We will, in this article, concentrate on the equilibrium properties of the membrane using Monte Carlo methods for mesh evolution. { Details pertaining to simulations can be found in literature,$^{\cite{Ho:1990p5747,Ho:1990p295,Lipowsky:1991p1059,Kroll:1992p968}}$ also see chapter by Gompper and Kroll$^{\cite{Piran:2003}}$ for an excellent review on  the application of triangulated surface techniques to various classes of problems.}\\ 
\section*{ In-plane  orientational order} \label{sec:inplane}
The  additional degrees of freedom, that arise from the anisotropy induced by the proteins, peptides, cytoskeletons and from tilt of lipids in the bilayer, can be   quantified by a $p$-atic vector living on the membrane surface. For example, the nematic order is defined as a $2\rm{-atic}$ vector field, with a $\pi$ rotational symmetry, in the local Darboux frame, constructed from the principal directions and the vertex normal at each vertex of the mesh. This means we need to calculate the principal curvatures and directions, unlike in Equation\eqref{eq:Hsurf} where the  squared mean curvature is directly approximated from the face normals. We will outline, in this article, the basic steps to compute these surface geometrical quantifiers.$^{\cite{Ramakrishnan:2010p1249}}$
\subsection*{Computing surface quantifiers}
From the neighborhood  of a vertex $v$, the normal to the surface at the vertex  is calculated as,
\[ \hat{N}(v)= \left[\sum_{\{f\}_v} \Omega[A(f)]\,\hat{N}(f) \right]/\left|\sum_{\{f\}_v}  \Omega[A(f)]\,\hat{N}(f)  \right|,\] where $A(f)$ is the area  and $\hat{N}(f)$  is the unit normal at face $f$. The weight factor $\Omega$ is chosen to be proportional to the area $A(f)$. The normals of the faces sharing an edge $e$, can be used to  approximate the normal to the edge as, \[ \hat{N}(e) = \left[\hat{N}[f_1(e)]+\hat{N}[f_2(e)]\right]/ \left| \hat{N}[f_1(e)]+\hat{N}[f_2(e)] \right|.\]

To quantify the curvature, that arises from the faces being non-planar, we define 
$H(e)=2 \left | \vec{R}(e) \right| \cos \left(\frac {\Phi(e)}{2} \right)$, along the direction \begin{math}\vec{B}(e)=\hat{N}(e) \times \hat{R}(e)\end{math}.$^{\cite{Hildebrandt:2004,Hildebrandt:2005p86}}$ $\Phi(e)$ is the signed dihedral angle between the faces, $f_{1}(e)$ and  $f_{2}(e)$, { taking a value of $\pi$ when the faces are coplanar} .  $\left | \vec{R}(e) \right| $ is the length of the edge. 
The discretized ``shape operator'',  that represents the curvature tensor and its orientation at $e$ is given by 
$\underline {\mathbf{S_e}}(e) = H(e) \left[ {\hat B}(e) \otimes {\hat B}(e) \right].$ It should be noted that $\underline {\mathbf{S_e}}(e)$ is defined at every edge containing vertex $v$, which can collectively be represented by  \{$\underline{\bf S_e}(e)$\}. These individual shape operators determine the curvature at $v$. The projection operator, $\underline{{\bf P}}(v) = \mathbbm{1} - \hat{N}(v)\otimes \hat{N}(v)$, projects $\underline {\mathbf{S_e}}(e)$  into the tangent plane at vertex $v$. Details of the surface quantifiers at $v$ are contained in the vertex shape operator, constructed as a weighted sum of these  projections, given by,
\begin{equation}
{\underline {\bf S_v}(v)} = \frac{1}{A(v)}\,\,\sum_{\{e\}_v} W(e)\,\underline{ {\bf P}}(v)^{\dagger}\,\underline{{\bf S_e}}(e)\,\underline{{\bf P}}(v).
\label{eq:curv_operator}
\end{equation}
$A(v)=\sum_{\{f\}_v} A(f)/3$ is the average surface area around $v$, while the weight factor for an edge is calculated as $W(e)= \hat{N}(v) \cdot \hat{N}(e)$. ${\underline {\bf S_v}(v)}$ is constructed in the global cartesian frame and a Householder transformation,$^{\cite{Taubin:1995,Ramakrishnan:2010p1249}}$ rotates the operator in Equation\eqref{eq:curv_operator} to its tangent  frame and the resulting matrix is a $2\times2$ minor. This transformation gives a computationally efficient route to calculate the eigenspectrum of Equation\eqref{eq:curv_operator}, when compared to using standard numerical techniques. The eigenvalues and eigendirections of ${\underline {\bf S_v}(v)}$ are respectively the principal curvatures($c_{1}$ \& $c_{2}$) and directions($\hat{e}_{1}$ \& $\hat{e}_{2}$). The scalar invariants are the mean curvature, $H=(c_{1}+c_{2})/2$ and the  Gaussian curvature, $R=c_{1}c_{2}$. We can now define the Darboux frame with basis vectors $[\hat{e}_{1}(v),\ \hat{e}_{2}(v),\hat{N}(v)]$ and the in-plane nematic orientation, in this frame, to be $\vec{m}=\cos\varphi(v)\hat{e}_{1}+\sin\varphi(v) \hat{e}_{2}$.  $\varphi(v)$ is the angle subtended by the nematic orientation with respect to maximum principal direction $\hat{e}_{1}$.
\subsection*{Parallel transport on a discrete mesh}
 In order to compare the orientation of two distant in-plane vectors on the surface, it is necessary to perform a parallel transport of the vectors on the discretized surface. In practice, we need only to define the parallel transport between neighboring vertices, i.e. a transformation  \begin{math}\hat{m} (v^{'}) \rightarrow \underline{\bf \Gamma}(v,\,v^{'}) \hat{m}(v) \end{math}, which brings $\hat{m}(v)$ correctly into the tangent plane of the vertex $v^{'}$, so that its angle with respect to the geodesic connecting $v$ and $v^{'}$ is preserved. 
If ${\hat r}(v,v^{'})$  is the unit vector  connecting   a vertex $v$ to its neighbor $v^{'}$ and ${\vec \zeta}(v)$=$\underline{{\bf P}}(v){\hat r}(v,v^{'})$  and ${\vec \zeta}(v^{'})$= $\underline{{\bf P}}(v^{'}) {\hat  r}(v^{'},v)$ are its projection on to the tangent planes at $v$ and $v^{'}$; then  our best estimate for  the directions of the geodesic  connecting them,  are the unit vectors ${\hat  \zeta}(v),\,{\hat \zeta}(v^{'})$. The decomposition of $\hat{m}(v)$ along the  orientation of the geodesic and its perpendicular in the tangent plane of $v$ is thus: 
\begin{equation}
 \hat{m}(v) =  \left\{\hat{m}(v) \cdot \hat{\zeta}(v)\right\}\hat{\zeta}(v)+ 
                    \left\{\hat{m}(v)\cdot [\hat{N}(v) \times \hat{\zeta}(v)]\right\} \left[\hat{N}(v) \times \hat{\zeta}(v)\right].
\end{equation} 
\noindent
Parallelism now demand that these  coordinates,  with respect to the geodesic orientation, are the same  in the tangent 
plane of $v^{'}$, therefore:
\begin{equation}
\underline{{\bf \Gamma}}(v,v^{'})\hat{m}(v)=  \left\{\hat{m}(v) \cdot \hat{\zeta}(v)\right\} \hat{\zeta}(v^{'})+ 
 \left\{\hat{m}(v) \cdot [\hat{N}(v) \times \hat{\zeta}(v)]\right\} \left[ \hat{N}(v^{'}) \times \hat{\zeta}(v^{'})\right].
\end{equation} 
\noindent
This parallel transport operation allow us to define the angle $\varphi({v,v^{'})}$ between vectors in the tangent plane at neighboring vertices, and in turn their cosine and sine as:
\begin{eqnarray}
 \cos(\varphi(v,v^{'}))&= & \hat{m}(v^{'})\cdot \underline{{\bf \Gamma}}(v,v^{'})\hat{m}(v) ;\\
 \sin(\varphi(v,v^{'}))&=& \left[\hat{N}(v^{'})\times\hat{m}(v^{'}) \right] \cdot \underline{{\bf \Gamma}}(v,v^{'})\hat{m}(v) \nonumber
\end{eqnarray}
\noindent
We can now define the lattice models with interaction between in-plane orientational field, e.g., the XY-model for a polar order($p=1$) on a random surface:
\begin{equation}
\label{eq:xy}
 {\cal H}_{1{\rm-atic}}= -\frac{\epsilon_{\rm 11}}{2}\sum_{\langle vv^{'} \rangle} \cos[\varphi(v,v^{'})]
\end{equation}
\noindent
or the Lebwohl-Lasher model for nematic order($p=2$) on a random surface:
\begin{equation}
\label{eq:leb}
 {\cal H}_{2{\rm-atic}}= -\frac{\epsilon_{\rm 22}}{2}\sum_{\langle vv^{'} \rangle}\left\{\frac{3}{2}\cos^2(\varphi(v,v^{'}))-\frac{1}{2}\right\}
\end{equation}
We will, in general, represent the self interaction energy of a $p$-atic field by $\mathscr{H}_{p-\rm{atic}}$.
%
\section*{Frustrated in-plane order and membrane morphology} \label{sec:res}
\begin{figure}
 \begin{center}
 \includegraphics[width=7.5cm]{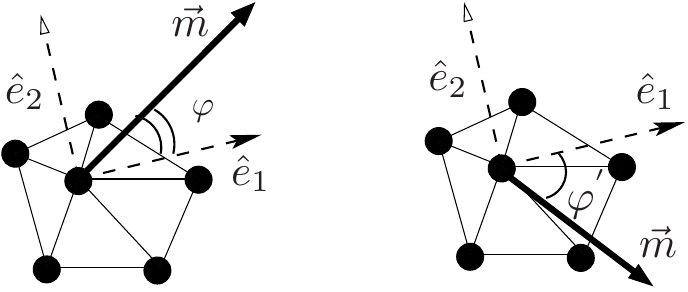}
 \end{center}
 \caption{\label{fig:fflip} A Monte Carlo move involving the in-plane field, $\vec{m}$. The membrane morphology remains unchanged while the field orientation is changed from $\varphi$ to $\varphi^{'}$. $\hat{e}_{1}$ and $\hat{e}_{2}$ are the principal directions.}
 \end{figure}

The curvature inducing properties of surface proteins, can be captured in a phenomenological model as an additional term to the RHS of  Equation\eqref{eq:Hsurf}, given by,
\begin{equation}
\label{eq:nem-mem}
\mathscr{H}_{NM}=\sum_{v} A(v) \left\{ \kappa_{\parallel}[H_{\parallel}(v)-c_{0}^{\parallel}]^2 \right. +
              \left. \kappa_{\perp}[H_{\perp}(v)-c_{0}^{\perp}]^{2} \right\}.
\end{equation}
 $\kappa_{\parallel}$  and  $\kappa_{\perp}$ are the directional bending rigidities parallel (or antiparallel)  and perpendicular to the  field $\vec{m}$. These phenomenological parameters, along with the directional spontaneous curvatures $c_{0}^{\parallel}$ and $c_{0}^{\perp}$, quantify the interaction strength of the protein with the membrane. The directional curvatures of the membrane at a vertex $v$, calculated using Gauss formula,$^{\cite{doCarmo:1976}}$ are  $H_{\parallel}(v)=c_{1}(v)\cos^{2}\varphi(v)+c_{2}(v)\sin^{2}\varphi(v)$ and $H_{\perp}=c_{2}(v)\cos^{2}\varphi(v)+c_{1}(v)\sin^{2}\varphi(v)$. The additional Monte Carlo move involving the in-plane order is given in Figure \ref{fig:fflip}.$^{\cite{Ramakrishnan:2010p1249}}$ \\

\begin{figure}[!h]
\begin{center}
\subfigure[]
{
\begin{minipage}[]{0.4\textwidth}
\vspace{0pt}
\includegraphics[width=5.75cm]{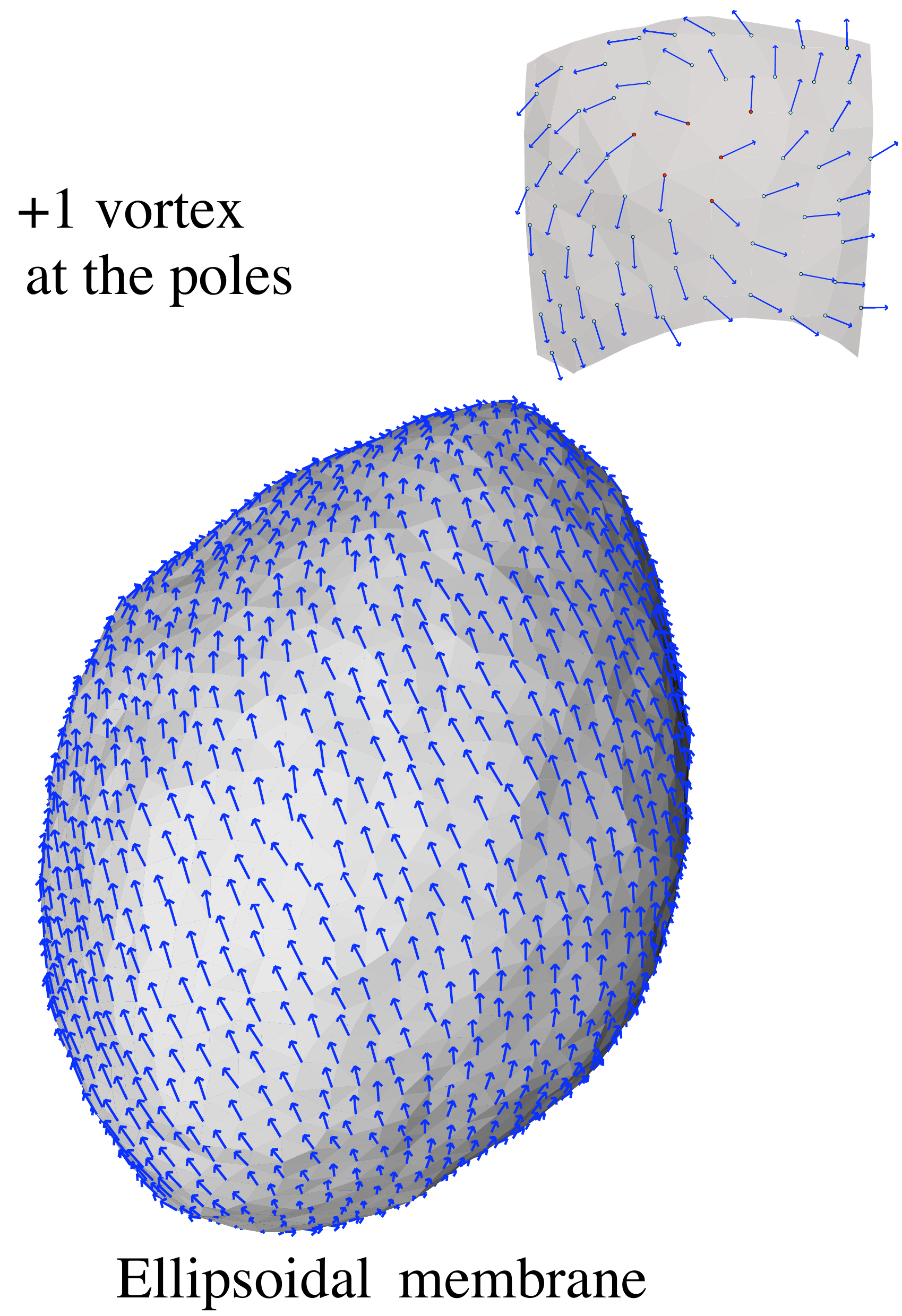}
\vspace{10pt}
\label{fig:xy-implicit}
\end{minipage}
}
\hspace*{40pt}
\subfigure[]{
\begin{minipage}[]{0.4\textwidth}
\vspace{0pt}
\includegraphics[width=5.75cm]{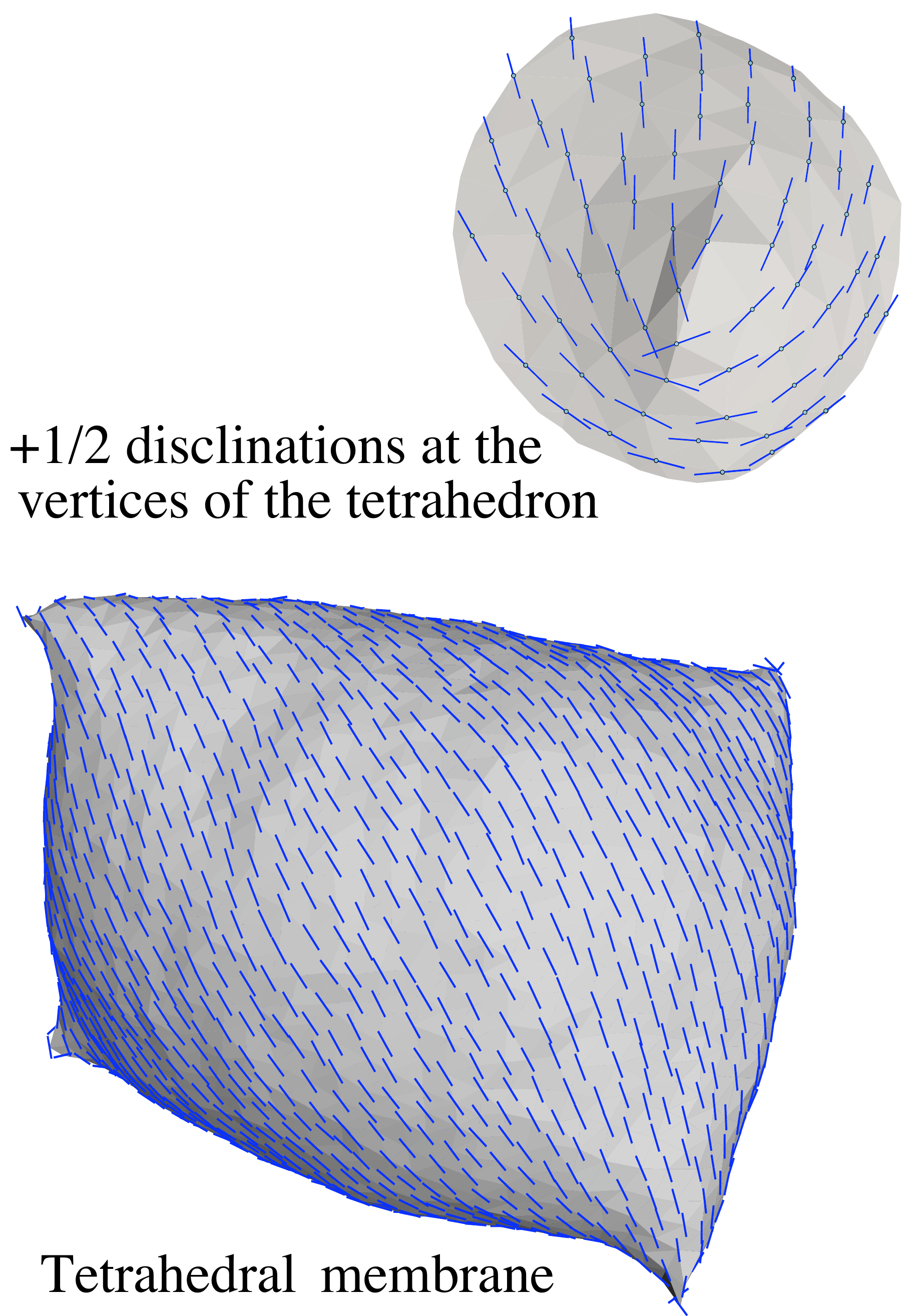}
\vspace{10pt}
\label{fig:ll-implicit}
\end{minipage}
}
\end{center}
\caption{ A surface of spherical topology with {\bf(a)} polar order ($p=1$, $\epsilon_{11}=10$) : Two +1 defects sit at the pole and an otherwise spherical membrane elongates into a cylinder  {\bf(b)} nematic order ($p=2$, $\epsilon_{22}=10$) : Four +1/2 disclinations remodel the surface into a tetrahedron. Simulations were performed with $\beta=1/k_{B}T=5.0$ and $\kappa=10$}
\end{figure}

The total energy of the field decorated surface, of spherical topology(genus, $g=0$), is $\mathscr{H}_{tot}=\mathscr{H}_{sur}+\mathscr{H}_{NM}+\mathscr{H}_{p-\rm{atic}}$. In the ground state of $\mathscr{H}_{tot}$, the $p$-atic  field is frustrated by the topology of the embedding surface and has in it $\chi p$ defects, each of topological charge $1/p$. In addition to the  bending rigidity, the equilibrium conformation of the surface now depends on the number and strength of the defects. This implicit coupling of membrane morphology to the in-plane order and the resulting membrane shapes has been  theoretically investigated in earlier works.$^{\cite{MacKintosh:1991p685,Lubensky:1992p531,Park:1992p946}}$ The equilibrium conformations of the membrane in our simulations, for $\kappa_{\parallel}=\kappa_{\perp}=0$, are in excellent agreement with these predictions. Figure \ref{fig:xy-implicit} shows the predicted ellipsoidal membrane for a polar field($p=1$), while the tetrahedral shape, characteristic of a nematic field, is shown in Figure \ref{fig:ll-implicit}. In case of the polar fields the two $+1$ antipodal vortices are positioned at the ellipsoidal caps whereas the four +1/2 disclinations are at the vertices of the tetrahedron for the nematic field. \\

\begin{figure}[!h]
\begin{center}
\subfigure[]{
\begin{minipage}[]{0.4\textwidth}
\vspace{3pt}
\includegraphics[width=5.75cm]{./xy-Drig0p25}
\vspace{10pt}
\label{fig:xy-0p25}
\end{minipage}
}
\hspace{30pt}
\subfigure[]{
\begin{minipage}[]{0.4\textwidth}
\vspace{3pt}
\includegraphics[width=5.75cm]{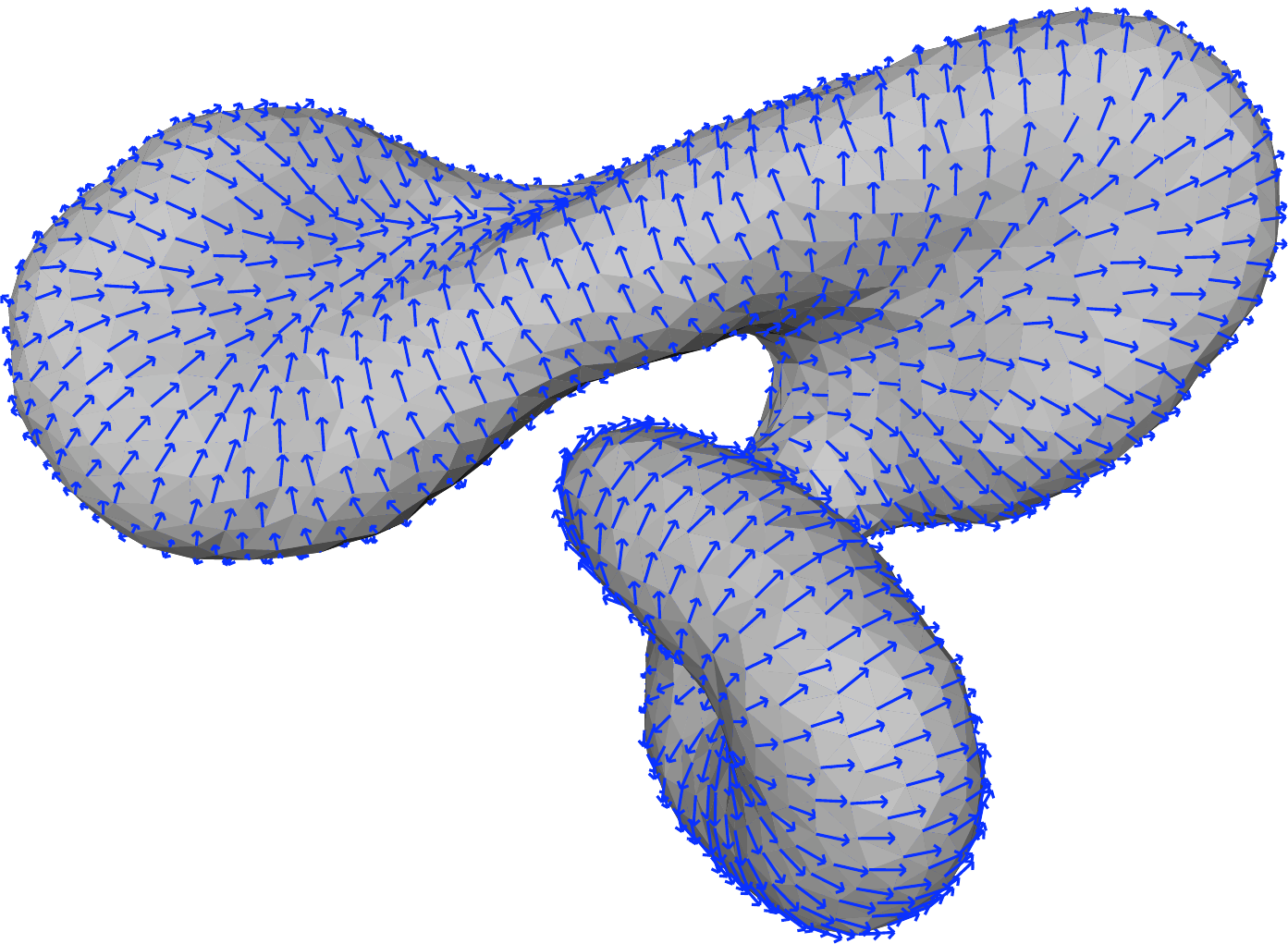}
\vspace{10pt}
\label{fig:xy-0p75}
\end{minipage}
}
\end{center}
\caption{\label{fig:nonzeroc0}Conformations of a fluid membrane with in-plane polar($p=1$) field for {\bf(a)} $C_{0}^{\parallel}=0.25$ and {\bf(b)} $C_{0}^{\parallel}=0.75$. Values of other parameters are $\kappa=10$, $\beta=5.0$, $\kappa_{\parallel}=5$ and $\epsilon_{11}=10$.}
\end{figure}

In the case of a fluid membranes, the shape change induced by the  ordered in-plane fields was investigated earlier by us.$^{\cite{Ramakrishnan:2010p1249}}$ It has been shown that the entropy dominated branched shapes, the equilibrium shapes of flexible surfaces($\kappa=0$), are cut off by order induced membrane stiffening. These observations add support to the hypothesis of membrane shape stabilization by protein-lipid interactions.\citep{McMahon:2005p274,Zimmerberg:2006p510} Further, non zero contributions from the directional rigidities and spontaneous curvatures drive the membrane into a wide array of shapes encompassing the biologically relevant {\it tubes, discs, caveolae} and {\it branches}.$^{\cite{Ramakrishnan:2011}}$ In this article,  using a polar in-plane field, we demonstrate the emergence of both axisymmetric and non axisymmetric membrane conformations as in Figure \ref{fig:nonzeroc0}. The polar field interaction with the membrane is thread like ($\kappa_{\perp}=0$) and hence deforms the membrane only along its long axis. \\

The {\it biconcave} and {\it twisted disc} shapes shown in Figure \ref{fig:xy-0p25} and \ref{fig:xy-0p75} are, respectively,  the ground state configurations for $C_{0}^{\parallel}=0.25$ and $C_{0}^{\parallel}=0.75$. The in-plane polar fields in Figure \ref{fig:nonzeroc0} have similar defect structures, two +1 vortices in this case, and hence have comparable energies of $\mathscr{H}_{1{\rm -atic}}$. Thus, the equilibrium membrane deformation is driven by the competition between the elastic energy (Equation \eqref{eq:Hsurf}) and field-membrane interaction energy(Equation \eqref{eq:nem-mem}). The minimum energy configuration of Equation\eqref{eq:Hsurf} is a minimal surface($H=0$) while that of Equation\eqref{eq:nem-mem} corresponds to a surface with the field oriented along the maximum principal direction $\hat{e}_{1}$, and the corresponding curvature $c_{1}=C_{0}^{\parallel}$.  The observed trend in the values of the  principal curvature, as a function of $C_{0}^{\parallel}$, are in agreement with the above arguments$^{\cite{Ramakrishnan:2011}}$.

\section*{Conclusion}
We have reviewed the basic techniques of dynamically triangulated surfaces used as model for fluid membranes with in-plane order in the macroscopic limit. The anisotropic membrane constituents are modeled as in-plane vector fields and surface rendering techniques suitable to study these systems have been introduced. The presence of an ordered in-plane $p$-atic field remodels the membrane into an ellipsoid($p=1$) and cylinder($p=2$). Introducing explicit field-membrane interactions widens the spectrum of the resulting shapes and can serve as an useful way of understanding the cooperative effect of protein lipid interactions. \bigskip \\

\noindent
Received: Jan 11, 2011 ; Revised: March 31, 2011 ;   DOI: 10.1002/mats.201100002 \medskip \\	 \bigskip \\
\noindent
Keywords: In-plane order ; lipid protein interactions ; membranes ; Monte Carlo simulation ; triangulated surfaces 


\end{document}